\documentclass[final,1p,times,twocolumn]{elsarticle}

\usepackage{amssymb}
\usepackage{amsmath}
\usepackage{color,soul}
\usepackage[usenames,dvipsnames]{xcolor}
\usepackage{graphicx}
\usepackage{xcolor, colortbl}
\usepackage{graphicx}
\usepackage{array}
\usepackage{tabularx}
\usepackage{booktabs}
\usepackage{makecell}

\newcommand{\tick}{\textcolor{ForestGreen}{\checkmark}}
\newcommand{\cross}{\textcolor{red}{\textbf{x}}}

\DeclareMathOperator{\softmax}{\mathrm{Softmax}}
\DeclareMathOperator*{\concat}{\mathrm{Concat}}
\DeclareMathOperator{\linear}{\mathrm{Linear}}

\journal{Computers in Biology and Medicine}

\begin{document}

\begin{frontmatter}

\title{Learning Hemodynamic Scalar Fields on Coronary Artery Meshes: A Benchmark of Geometric Deep Learning Models}

\author[polimi]{Guido Nannini}
\author[twente]{Julian Suk}
\author[twente]{Patryk Rygiel}
\author[aumc,iiua,polimi]{Simone Saitta}
\author[polimi]{Luca Mariani}

\author[monz]{Riccardo Maranga}
\author[monz,umiclin]{Andrea Baggiano}
\author[monz,umibio]{Gianluca Pontone}

\author[twente]{Jelmer M. Wolterink}
\author[polimi]{Alberto Redaelli}

\affiliation[polimi]{organization={Department of Electronics Information and Bioengineering, Politecnico di Milano},
            city={Milan},
            country={Italy}}
\affiliation[twente]{organization={Department of Applied Mathematics and Technical Medical Center, University of Twente},
            city={Enschede},
            country={The Netherlands}}
\affiliation[monz]{organization={Department of Perioperative Cardiology and Cardiovascular Imaging, Centro Cardiologico Monzino IRCCS},
            city={Milan},
            country={Italy}}
\affiliation[umibio]{organization={Department of Biomedical, Surgical and Dental Sciences, University of Milan},
            city={Milan},
            country={Italy}}
\affiliation[umiclin]{organization={Department of Clinical Sciences and Community Health, University of Milan},
            city={Milan},
            country={Italy}}
\affiliation[aumc]{organization={Department of Biomedical Engineering and Physics, Amsterdam UMC},
            city={Amsterdam},
            country={The Netherlands},
            }
\affiliation[iiua]{organization={Informatics Institute, University of Amsterdam},
            city={Amsterdam},
            country={The Netherlands},
            }

\begin{abstract}
Coronary artery disease involves the narrowing of coronary vessels due to atherosclerosis and is currently the leading cause of death worldwide. The gold standard for its diagnosis is the fractional flow reserve (FFR) examination, which measures the trans-stenotic pressure ratio during maximal vasodilation. However, the invasiveness and cost of this procedure have prompted the development of computer-based virtual FFR (vFFR) computation, which simulates coronary flow using computational fluid dynamics (CFD) techniques.
Geometric deep learning algorithms have recently shown the capability to learn features on meshes, including applications in cardiovascular research. In this work, we aim to conduct a comprehensive empirical analysis of different backends for predicting vFFR fields in coronary arteries, serving as surrogates for CFD simulations. We evaluate six different backends and compare their performance in learning hemodynamics on meshes using CFD solutions as ground truth.
This study is divided into two main parts:
i) First, we use a dataset of 1,500 synthetic bifurcations of the left coronary artery. Each model is trained to predict various pressure-related fields, from which the vFFR field is reconstructed. We compare the models’ performance when different learning variables are used during training.
ii) Second, we use a dataset of 427 patient-specific CFD simulations from a previous study by our group. Here, we repeat the experiments conducted on the synthetic dataset, focusing on the learning variable that yielded the best performance in the synthetic dataset.
Most backends achieved very good performance on the synthetic dataset, particularly when learning the pressure drop over the manifold. For other network output variables (e.g., pressure and the vFFR field), transformer-based backends outperformed all other architectures. When trained on patient-specific data, transformer-based architectures were the only ones to achieve strong performance, both in terms of average per-point error and in accurately predicting vFFR in stenotic lesions.
Our findings indicate that various geometric deep learning backends can serve as effective CFD surrogates for problems involving simple geometries. However, for tasks involving datasets with complex and heterogeneous topologies, transformer-based networks are the optimal choice. Additionally, pressure drop emerged as the optimal network output for learning pressure-related fields.
\end{abstract}

\begin{graphicalabstract}
\includegraphics[width=\textwidth]{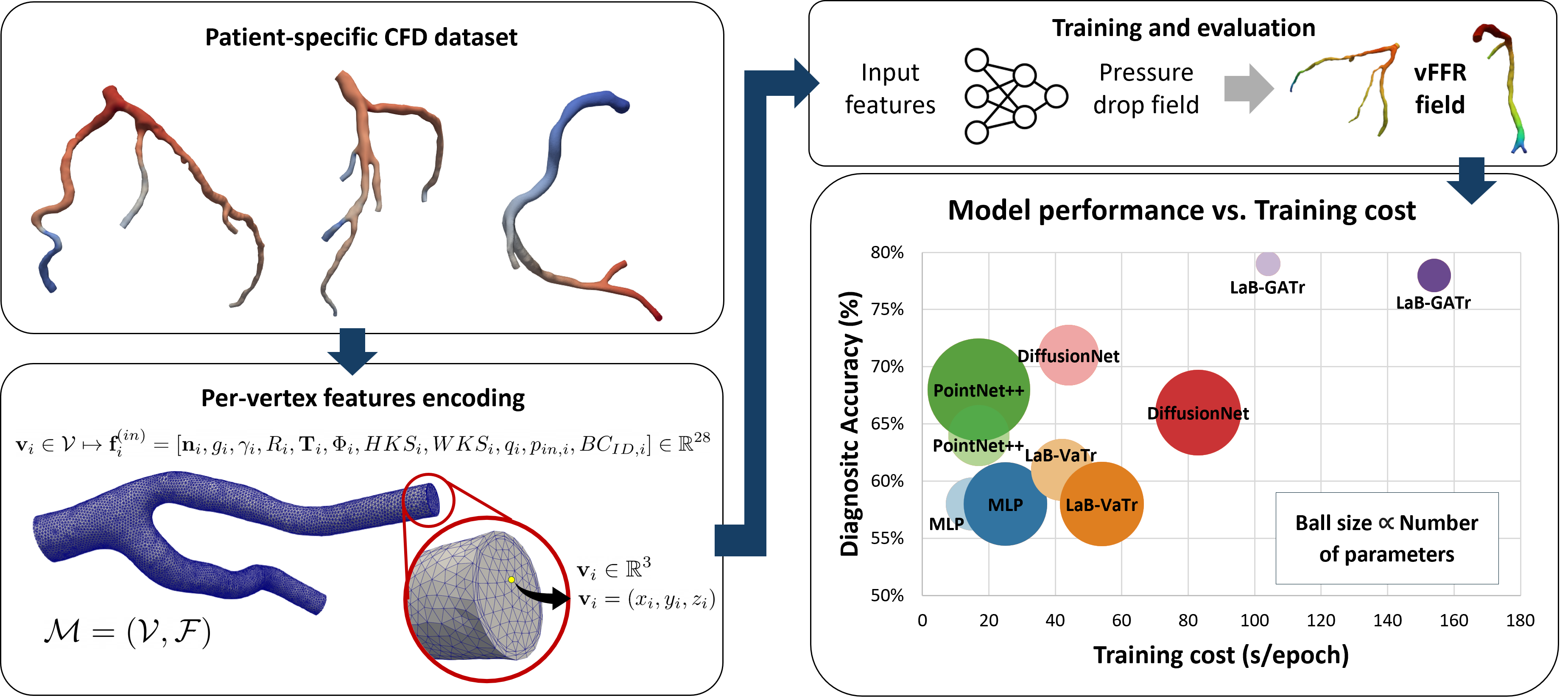}
\end{graphicalabstract}

\begin{highlights}
\item Various geometric deep learning techniques are suitable as CFD surrogates for problems involving simple meshes and topologies, with transformer-based models outperforming other methods.
\item Transformer-based architectures were found to be the most suitable models to generalize on complex shapes such as patient-specific data.
\item When predicting pressure-related scalar fields defined over meshes of (coronary) arteries, using pressure drop as a learning variable during the training is preferable, as it yields better performance compared to learning pressure or FFR field directly.
\item Learning pressure-related fields on artery meshes requires essential knowledge of the surface's geometric features and physical boundary conditions. Providing the pressure at the mesh inlet significantly enhances model performance.
\end{highlights}


\end{frontmatter}

\section{Introduction} \label{intro}

Coronary artery disease (CAD) is the leading cause of death worldwide \cite{okrainec2004coronary}. It consists of the buildup of fibro-lipidic and calcific plaque in the walls of the coronary arteries, leading to narrowing (i.e., stenosis) of the vessel lumen and reduced blood flow. Ultimately, the limited supply of oxygen can result in myocardial infarction and death \cite{libby2005pathophysiology}. In recent years, fractional flow reserve (FFR), defined as the ratio between the pressure downstream of a stenosis and the pressure at the coronary ostium, has emerged as the gold standard to assess the functional severity of a stenosis \cite{pijls1995fractional}. FFR is invasively measured \textit{in-vivo} with a pressure probe during maximal hyperemia induced by vasodilator administration. Stenting is recommended when $\text{FFR}<0.8$, and medical treatment otherwise \cite{pijls2012functional}. Several clinical trials have shown that invasive FFR is a more robust criterion for identifying potential culprit lesions compared to stenosis degree alone \cite{van2015fractional, koo2011diagnosis, norgaard2014diagnostic}. However, its use in clinical practice remains limited due to the invasiveness and cost of the procedure \cite{tonino2009fractional, van2015fractional, zimmermann2015deferral}.

Non-invasive computation of FFR requires knowledge of the pressure field within the coronary volume and along the vessel walls. Computational fluid dynamics (CFD) enables \textit{in-silico} simulation of blood flow within the lumen of coronary arteries, typically reconstructed from computed tomography (CT). CFD results can be used to obtain a non-invasive virtual measurement of FFR, known as vFFR \cite{taylor2013computational, coenen2015fractional, zarins2013computed}. Although the primary advantage of vFFR is its ability to overcome the invasiveness of FFR measurement, a full CFD simulation of coronary flow over a single cardiac cycle can be highly computationally demanding ($>24$ hours) and requires expert knowledge throughout the process \cite{candreva2022current, zhong2018application}. 
Different strategies to speed up the computation of vFFR have been proposed in the literature, such as using reduced order modeling \cite{bezerra2018tct, boileau2018estimating, itu2012patient} and hybrid 3D-1D models \cite{grande20221d}. While being computationally less expensive, such models fail to capture the complexity of 3D flows, potentially introducing errors in estimating the pressure and velocity field \cite{pfaller2022automated}. Additionally, parameters such as time and space discretization, as well as boundary conditions (BCs), are subject to inter-operator variability, introducing uncertainty in the solution of the CFD simulation and making the overall process challenging to standardize \cite{sankaran2016uncertainty, valen2018real}.
In a previous study by our group \cite{nannini2024automated}, we proposed a CFD pipeline that automatically tunes the BCs while computing the solution, partially solving the inter-operator bias. Furthermore, we demonstrated that vFFR can be accurately computed by simulating coronary flow in a steady state with BCs averaged over the cardiac cycle, yielding good diagnostic accuracy ($\geq$90\%), comparable to that of pulsatile simulations with time-varying BCs. Steady-state CFD preserves the 3D flow development while significantly reducing computational costs, although it still requires 30 to 60 minutes per patient.

Data-driven models allow an additional step forward to accelerate the simulation process. In cardiovascular modeling, deep learning has been used to automate different aspects of the CFD pipeline, such as image segmentation and 3D model reconstruction \cite{nannini2024fully, gu2021fusing, gharleghi2022automated, saitta2022deep, alblas2023sire}. However, solving the physical problem (e.g., the Navier-Stokes equations), which represents the most computationally demanding step, is still typically reliant on numerical algorithms. Hence, (partial) replacement of this step by data-driven deep learning methods is an active research domain. 


Machine learning models can learn to generalize hemodynamics from a training dataset to unseen data based on geometric (e.g., the anatomy) and physical (e.g., the BCs) features. One approach to this involves parametrizing the lumen centerline or the mesh of the arterial wall in a 1D or 2D Euclidean domain and using fully connected networks, such as multilayer perceptrons (MLPs) or convolutional neural networks (CNNs), to estimate scalar or vector fields based on shape descriptors
\cite{itu2016machine, su2020generating}. 

Alternatively, models can be developed that operate directly on the original artery geometry without requiring a mapping to a lower-dimensional Euclidean domain. Such models typically fall in the realm of geometric deep learning (GDL), i.e., deep learning models for data in non-Euclidean domains, such as graphs, meshes, and point clouds \cite{bronstein2017geometric, bronstein2021geometric}. For example, Rygiel \textit{et al.} \cite{rygiel2023centerlinepointnet++} proposed a variant to PointNet++ for predicting the pressure drop on a dataset of synthetic anatomies of both left and right coronary arteries. Wang \textit{et al.} \cite{wang2023deep} trained a point cloud network (PCN) to infer the pressure and velocity field in the carotid artery and in the aorta; Zhang \textit{et al.} \cite{zhang2023physics} used a physics-informed variant of the PointNet to predict the 3D velocity field in the abdominal aorta; Suk \textit{et al.} \cite{suk2023se} used a multiscale E(3)-steerable graph neural network (GNN) for estimating the velocity field in synthetic coronary arteries. Suk \textit{et al.} \cite{suk2024lab} presented a variant to the geometric algebra transformer (GATr) model \cite{brehmer2024geometric}, specifically for large biomedical meshes (LaB-GATr), achieving very good wall shear stress (WSS) and velocity estimation in large synthetic coronary arteries models. Pegolotti \textit{et al.} \cite{pegolotti2024learning} proposed a data-driven reduced order model based on a GNN that learns 1D quantities along the centerline of the vascular tree. Likewise, Gharleghi \textit{et al.} \cite{gharleghi2022transient} proposed a method to map the original surface to a fixed 2D Euclidean domain and applying CNN multiscale-neural network for WSS inference on the surface of the bifurcation of the left coronary artery. The main drawback of such techniques is that parametrization may not be feasible for complex shapes (e.g., in the presence of pathology), limiting their applicability.
While demonstrating the potential of GDL techniques for predicting hemodynamic quantities on and in meshes, most studies are conducted on synthetic or simplified geometries due to the commonly limited availability of large patient-specific datasets. This makes it difficult to identify common patterns and answer questions about the right data representation and the most appropriate model architecture and size. 

In this work, we address these questions for the problem of estimating hemodynamic fields on the surface of real-life patients' coronary artery 3D anatomies. Specifically, we focus on the problem of learning the scalar pressure field on a surface from which we compute the corresponding FFR field. We perform an extensive empirical comparison of architectures that are different in nature and either well-established or increasingly used in biomedical applications and beyond. These are the multi-layer perception (MLP); PointNet++ \cite{qi2017pointnet}; DiffusionNet \cite{sharp2022diffusionnet}; DeltaConv \cite{wiersma2022deltaconv}; and two transformer models, one exploiting geometric algebra representations (i.e., LaB-GATr) and the other, referred to as \textit{\textbf{va}nilla} transformer, operating under linear algebra (i.e., LaB-VaTr) \cite{suk2024lab}. We train and validate these models using results from steady CFD simulations on two datasets: a synthetic set of 1500 anatomies representing the left coronary artery bifurcation and a patient-specific set of 427 cases validated against invasive FFR measurements. We train our models using both geometric and physical input features, representing the 3D anatomy and the BCs of the corresponding CFD simulation. Moreover, we compare the diagnostic accuracy of the methods.

The main contributions of this paper are the following: 
(\textit{i}) We demonstrate the feasibility of leveraging GDL to predict vFFR scalar fields in real-world patient cases; (\textit{ii}) We systematically compare the performance of various architectures and model sizes for inferring scalar fields on the surface of coronary arteries, finding the superior performance of transformer models in this task; 
(\textit{iii}) We train our models to learn various pressure-related variables, demonstrating that predicting the pressure drop is optimal for deriving both the pressure and vFFR fields.

\section{Methods} \label{methods}

\begin{figure}[h]
\includegraphics[width=10cm]{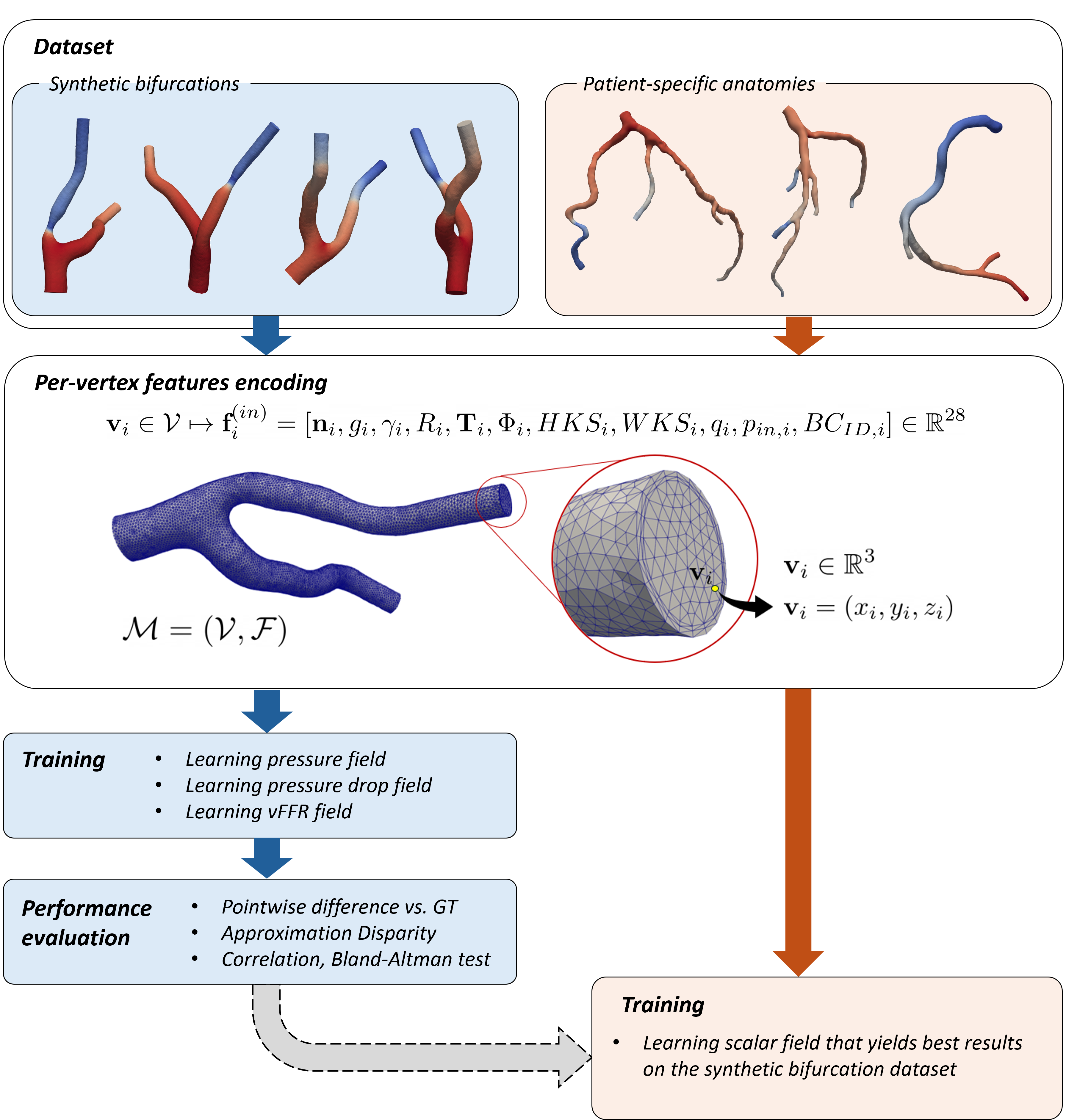}
\centering
\caption{Representation of the workflow adopted for the training of all models. First, models are trained on the synthetic dataset to learn one of three output variables; after evaluation, they are retrained on the patient-specific dataset using the best-performing variable, with per-vertex feature encoding applied consistently across both datasets. Blue boxes and arrows refer to the pipeline applied to synthetic datasets; orange ones refer to patient-specific data.}
\label{fig:workflow}
\end{figure}

The aim of this study was to systematically compare different models for learning physics-based pressure-related fields on meshes of coronary arteries. Figure \ref{fig:workflow} provides a schematic representation of our systematic approach in two different data sets. We first performed experiments using a synthetically generated dataset of geometries representing the left coronary bifurcation. Since our primary interest is in predicting FFR, we trained each model to learn one of three related scalar fields: (\textit{i}) pressure; (\textit{ii}) pressure drop, and (\textit{iii}) the vFFR field. Based on the results obtained in the synthetic dataset, we then selected one output variable among these three for further experiments in patient-specific anatomies. For a fair comparison between models, all models were fed the same vertex-specific input features. Under the assumption of laminar and developed flow, the pressure field (and consequently the pressure drop and the FFR) can be considered constant on the cross-section of the vessel, thus fully represented by the value assumed at the wall. Thus, we consider the pressure to be constant across the vessel cross-section, allowing us to train our models exclusively on the surface mesh while excluding internal nodes of the volumetric mesh. This approach significantly reduces the computational cost both in training and inference.

\subsection{Data}

\subsubsection{Synthetic dataset}\label{sec_synt}
We generated a dataset of 1,500 anatomies representing the bifurcation of the left coronary artery into the left anterior descending and circumflex arteries. To create this, we used a bifurcation model reconstructed from a healthy subject, capturing approximately 1 cm upstream and 4 cm downstream of the bifurcation. Randomly positioned stenoses (ranging from 1 to 3 per model) were introduced with severity levels representing caliber reductions between 20\% and 60\%, including both eccentric and concentric stenoses. Each synthetic model was discretized into a tetrahedral volumetric mesh, then a CFD simulation was conducted in SimVascular \cite{updegrove2017simvascular}, setting boundary conditions randomly sampled within physiological ranges to achieve an inflow $Q \sim \mathcal{U}(100, 300)$ ml/min and an aortic pressure $p_{ao} \sim \mathcal{U}(90, 110)$ mmHg. Flow was assumed to be laminar and incompressible, with density $\rho = 1.05$ g/cm\textsuperscript{3} and dynamic viscosity $\mu = 4$ cP.

\subsubsection{Patient-specific dataset}
To test the models on real anatomies, we used data from patient-specific CFD simulations generated in a previous study by our group \cite{nannini2024automated}. Briefly, 105 geometries of the left ($n=76$) and right ($n=29$) coronary arteries were reconstructed from CT scans acquired at Centro Cardiologico Monzino (Milan, Italy) for patients scheduled for clinically indicated invasive coronary angiography due to suspected CAD. Segmentations were performed in 3DSlicer \cite{fedorov20123dslicer}, including the coronary artery volume downstream of the ostium to where the vessel diameter decreased below 1.5 mm, following established guidelines \cite{pontone2018training}. The reconstructed anatomies were then discretized into tetrahedral elements with a characteristic size of 0.25 mm using the TetGen algorithm within SimVascular \cite{updegrove2017simvascular}. For each mesh, 3 to 5 steady CFD simulations were run, slightly varying the BCs estimated from baseline clinical measurements (i.e., the average hyperemic inflow $\bar{Q}_{hyp}$ and the outlet resistances $R_{out}$), to match the patient's aortic pressure. For further details on the CFD simulations, see Ref. \cite{nannini2024automated}. A total of 427 CFD simulations were conducted, with the results serving as ground truth for training the networks. The present study was performed in accordance with recommendations of the local Ethics Committee, with written informed consent from all subjects, in accordance with the Declaration of Helsinki.

\subsection{Network architectures}

In this work, we considered six different deep learning architectures, namely a simple, fully connected network (MLP), a point cloud network (PointNet++), two mesh neural network, leveraging approximated local convolutions in the neighborhood of each node (DiffusionNet, DeltaConv), and two transformer models (LaB-VaTr, LaB-GATr). For each model, we defined a small (S) version and a large (L) version to mitigate any bias related to model size: each S-model had approximately 1M parameters, while each L-model contained 3M parameters. An exception to this is LaB-GATr, which only had 230k and 425k parameters for the S and L versions, respectively, -justified by its task-specific SO(3) equivariant layers and the runtime overhead. A brief description of each architecture, highlighting its main features, is provided below, and the characteristics of the models trained in this study are reported in Table \ref{tab:model_variants}. All networks were implemented in Python, using the PyTorch \cite{paszke2019pytorch} and PyTorch Geometric \cite{fey2019fast} libraries.

\begin{table}[ht]
\centering
\caption{Key characteristics of the model variants used in this study. Depth represents the number of building blocks in each network (e.g., layers for MLP, hierarchies for PointNet++). Channels denote the number of latent channels. Downsampling factors (NA=Not Applicable) specify the compression applied to the mesh prior to input into the network.}
\begin{tabular}{l|l|c|c|c}
\hline
\textbf{Model}    & \textbf{Variant}  & \textbf{Depth} & \textbf{Channels} & \textbf{Downsampling factors}    \\ \hline
MLP               & S variant         & 6              & 512               & NA                                \\ 
                  & L variant         & 12             & 512               & NA                                \\ \hline
PointNet++ \cite{qi2017pointnet}       & S variant         & 4              & 256               & 0.05; 0.50; 0.75; 0.90           \\ 
                  & L variant         & 4              & 512               & 0.05; 0.50; 0.75; 0.90           \\ \hline
DiffusionNet \cite{sharp2022diffusionnet}      & S variant         & 3              & 256               & NA                                \\ 
                  & L variant         & 6              & 256               & NA                                \\ \hline
DeltaConv \cite{wiersma2022deltaconv}         & S variant         & 8              & 128               & 0.10                              \\ 
                  & L variant         & 8              & 256               & 0.10                              \\ \hline
LaB-VaTr \cite{suk2024lab}          & S variant         & 12             & NA                & 0.05                              \\ 
                  & L variant         & 24             & NA                & 0.05                              \\ \hline
LaB-GATr \cite{suk2024lab}          & S variant         & 8              & 8                 & 0.05                              \\ 
                  & L variant         & 16             & 8                 & 0.05                              \\ \hline
\end{tabular}
\label{tab:model_variants}
\end{table}

\subsubsection{Multi-layer Perceptron (MLP)}

We included as a baseline model a point-wise MLP, which consisted of fully connected layers with batch normalization \cite{ioffe2015batch}. We used as activation function the Leaky ReLU \cite{xu2015empirical}, with a negative slope of 0.2. Throughout our work, we denote MLPs with $h_{\theta}$.

\subsubsection{PointNet++}

PointNet++ \cite{qi2017pointnet} is a popular point cloud network that applies a per-point MLP followed by max aggregation over the $k$-nn neighborhood $\mathcal{B}(i)$ of the $i$-th vertex of the mesh. The message-passing layers are defined by the following equation:

\begin{equation}
    \mathbf{x}_i^{(l+1)} = \max_{\ j \in \mathcal{B}(i) \ } h_{\theta} (\mathbf{x}_j^{(l)}, \mathbf{r}_{i-j})
    \label{mp_pointnet}
\end{equation}

where $\mathbf{x}_i^{(l)}$ are the features of the $i$-th vertex in layer $l$, $h_{\theta}$ is an MLP of arbitrary depth, and $\mathbf{r}_{i-j}$ represents the relative position between vertices. PointNet++ uses sampling and grouping operations to hierarchically subsample mesh vertices in the contracting pathway and perform interpolation in the expanding pathway. To handle the large scale of the mesh, we utilized a network with four hierarchical levels, employing cumulative sampling factors of 0.05, 0.50, 0.75, and 0.90.

\subsubsection{DiffusionNet}

DiffusionNet \cite{sharp2022diffusionnet} approximates local convolutions to learn through the learned feature diffusion process by leveraging the eigendecomposition of the Laplacian operator and applying spatial gradients. The network consists of multiple blocks, where an MLP is applied to each point to represent pointwise features, learned diffusion (for spatial communication), and spatial gradient features (to enable directional filters). The diffusion layer $d_t(\mathbf{x})$ is evaluated using the precomputed eigenvectors $\Phi:=[\phi_i]$ and eigenvalues $\lambda_i$ of the Laplace-Beltrami operator (denoted as $\Delta$):

$$
    h_t(\mathbf{x}) = \Phi_{\Delta} \begin{bmatrix} e^{-\lambda_0t} \\ e^{-\lambda_1t} \\ \hdots \end{bmatrix} \odot (\Phi^T_{\Delta} M \mathbf{x})
$$

where $M$ denotes the mass matrix, $\odot$ the elementwise product and $t$ is the learned diffusion time. This closed-form solution to the heat equation in eigenvector basis is known as spectral acceleration~\cite{sharp2022diffusionnet}. Spatial gradient features are obtained by learning a $D \times D$ matrix $A$ (with $D$ equal to the number of features) to compute the spatial gradient $\text{grad}_A(\mathbf{x})$. Features propagation through DiffusionNet blocks is then defined by:

\begin{equation}
    \mathbf{x}_i^{(l+1)} = h_{\theta} 
    \left( 
    d_t(\mathbf{x}_i^{(l)}),
    \text{grad}_A(\mathbf{x}_i^{(l)}),
    \mathbf{x}_i^{(l)}
    \right)
    \label{mp_diffusion}
\end{equation}

\subsubsection{DeltaConv}

DeltaConv \cite{wiersma2022deltaconv} exploits anisotropic and intrinsic convolutional layers to learn on surfaces. DeltaConv utilizes two streams for feature propagation: a scalar feature stream ($\mathbf{x}$) and a vector feature stream ($\mathbf{u}$). These streams exchange information through geometric operators that convert scalar features to vector features and vice versa. Message passing in DeltaConv is defined as follows:

\begin{subequations} \label{mp_deltaconv}
\begin{align}
    \mathbf{u}_i^{(l+1)} &= h_{\theta_0} 
    \left(
    \mathbf{u}_i^{(l)}, \text{grad} \, \mathbf{x}_i^{(l)}, \Delta \mathbf{u}_i^{(l)}
    \right) \label{mp_deltaconv_a} \\
    \mathbf{x}_i^{(l+1)} &= h_{\theta_1} 
    \left(
    \mathbf{x}_j^{(l)},
    \text{div} \, \mathbf{u}_i^{(l+1)},
    \text{curl} \, \mathbf{u}_i^{(l+1)},
    ||\mathbf{u}_i^{(l+1)}||
    \right) 
    + \max_{j \in \mathcal{B}(i)} h_{\theta_2} (\mathbf{x}_j^{(l)})
    \label{mp_deltaconv_b}
\end{align}
\end{subequations}

Similar to PointNet++, an MLP is applied point-wisely, followed by maximum aggregation over the neighborhood $\mathcal{B}(i)$ of each vertex in the mesh. To account for the large scale of the data, the mesh was downsampled using farthest point sampling before being input into the network. In our implementation, we set a downsampling ratio of 0.1 and defined $\mathcal{B}(i)$ to include 30 neighbors.

\subsubsection{LaB-GATr and LaB-VaTr}
Large biomedical geometric algebra transformers (LaB-GATr)~\cite{suk2024lab} are scalable self-attention models. LaB-GATr applies compression via cross-attention from the source resolution to a coarse token set and learned interpolation back to the original resolution. Across the coarse token set, feature vectors are updated through multi-head attention blocks:
\begin{gather*}
    a_t = \softmax\left( \frac{q_t\left( \mathbf{x}_i^{(l)} \right) k_t\left( \mathbf{x}_i^{(l)} \right)^T}{\kappa} \right) v_t\left( \mathbf{x}_i^{(l)} \right)\\
    A = \mathbf{x}_i^{(l)} + \linear\left( \concat_t a_t \right)\\
    \mathbf{x}_i^{(l+1)} = A + h_{\theta}(A)
\end{gather*}
where queries $q$, keys $k$ and values $v$ are linear layers, $\kappa \in \mathbb{R}$ is a scaling factor and attention heads are indexed by $t$. LaB-GATr operates in the projective geometric (Clifford) algebra $\mathbf{G}(3, 0, 1)$ which requires expressing $\mathbf{x}_i^{(l)}$ as so-called multivectors. For further details, refer to \cite{brehmer2024geometric}. The advantage of the attention framework is its global context aggregation after only one block, which facilitates learning on large meshes. To make training the model computationally feasible, we applied a compression factor of 0.05 to downsample the input mesh.

We also included in our benchmark LaB-VaTr, a variant of the LaB-GATr architecture that retains the same layout but replaces geometric algebra with \textit{\textbf{va}nilla} fully connected layers in linear algebra.

\subsection{Equivariance}\label{sec:equiv}
All networks considered in this study exhibit permutation-equivariance. This means that if the input points are permuted (through a permutation $P$), the output at each stage of the network is permuted in the same way. Denote networks as $f_\theta$. This property can be expressed by the following condition:
$$
(f_\theta \circ P)(\cdot) \equiv (P \circ f_\theta)(\cdot), \quad \forall P \in S_n,
$$
where \(\circ\) denotes composition, and \(S_n\) represents the symmetric group of permutations on \(n\) elements. Message passing and self-attention mechanisms are examples of permutation-equivariant building blocks commonly used in geometric deep-learning architectures. 

In addition to permutation equivariance, the LaB-GATr is SE(3)-equivariant, meaning that if rotation and translation are applied to the input mesh, through a group action $\varrho$, the network will rotate and translate its output in the same manner, preserving the relationship between the mesh and its transformed state:
$$
(f_\theta \circ \varrho)(\cdot) \equiv (\varrho \circ f_\theta)(\cdot), \quad \forall \varrho \in \text{SE}(3),
$$
Furthermore, DeltaConv is invariant to, i.e., unaffected by, any orientation-preserving isometry such as roto-translation.

\subsection{Experiments}

First, we trained each model on the synthetic dataset described in Section \ref{sec_synt}. The dataset was divided into 1,100 training cases, 200 validation cases, and 200 test cases. All models were trained for 500 epochs with a batch size of 2 on NVIDIA L40 (48 GB) GPUs, using the Adam optimizer with an initial learning rate of $3 \cdot 10^{-4}$ and an exponential decay rate of $\gamma_{decay} = 0.987$. We performed three training runs using different target variables, i.e., pressure, pressure drop, and FFR (see Section \ref{sec_output}). For the patient-specific dataset, we only trained models using pressure drop as a target variable. To improve the stability of the training process, all target variables $y$ were standardized using the formula $y_{std,i} = (y_i - \mu_y) / \sigma_y$, where $\mu_y$ and $\sigma_y$ represent the mean and standard deviation of $y$ across the training dataset. All models were trained using the mean squared error (MSE) as the loss function.

\subsubsection{Input features}

Input features and their definitions are summarized in Table \ref{tab:features}. Let $\mathcal{M}=(\mathcal{V}, \mathcal{F})$ represent surface mesh of vertices $\mathcal{V}$ and faces $\mathcal{F}$. We defined the input features by assigning each vertex $\mathbf{v} \in \mathcal{V}$ a set of $c_{in}$ features, $f^{in}: \mathcal{V} \rightarrow \mathbb{R}^{c_{in}}$. These features capture both local and global morphological properties of the mesh, as well as physical parameters describing the BCs of the CFD simulation. 
For each anatomy in the dataset, the following steps were carried out to construct the set of input features. For each mesh vertex, we computed the outward surface normal and appended its components as features. Next, the vessel centerline was reconstructed using the Vascular Modeling Toolkit (VMTK) library \cite{piccinelli2009framework}. Geometric properties calculated along the centerline (e.g., the radius of the maximum inscribed sphere) were then projected onto the surface by matching each surface point to the nearest point on the centerline. The radius, the curvilinear abscissa, and the projection of the tangent to the centerline onto the surface, approximating the parallel transport field, were appended as input features. The tangent direction to the centerline has previously been used as an input feature for GNNs trained to learn hemodynamic fields, as it serves as an indicator of flow direction through the fluid volume \cite{pegolotti2024learning}. Next, we appended the shortest geodesic distance from each vertex of the mesh to the inlet surface, which we computed as described in \cite{nannini2024fully}. We included three local shape descriptors derived from the spectral decomposition of the Laplace-Beltrami operator: the Laplacian eigenvectors \cite{levy2006laplace}; the heat kernel signature (HKS) \cite{sun2009concise}, which characterizes heat diffusion on a surface; and the wave kernel signature (WKS) \cite{aubry2011wave}, based on the Schrödinger wave equation. These descriptors provide a pointwise characterization of each vertex in relation to the entire mesh. We computed the discretized signature matrices using the Laplace eigenfunctions up to the 5\textsuperscript{th} order and appended the results to the input features.
Finally, we added constant scalar fields over $\mathcal{V}$ to represent the global parameters of the CFD simulation BCs, including blood inflow and inlet pressure. We remark that, while the pressure at the inlet surface (i.e., the coronary ostium) is generally a result of the CFD analysis, thus not known \textit{a priori}, it can be safely approximated with the aortic pressure, which is typically known for patients.
We also assigned a constant value from the set [0, 1, 2] to each vertex to indicate its corresponding node type in the CFD analysis: [wall, inlet, outlet], respectively. 
In summary, we associated each vertex of the mesh $\mathbf{v}_i$ to the vector of features $\textbf{f}_{i}^{(in)}$:
$$
\textbf{v}_{i} \in \mathcal{V} \mapsto \textbf{f}_{i}^{(in)} = [\textbf{n}_i, g_i, \gamma_i, R_i, \textbf{T}_i, \Phi_i, HKS_i, WKS_i, q_i, p_{in,i}, BC_{ID,i} ] \in \mathbb{R}^{28}
$$

\begin{table}[h!]
    \centering
    \caption{Input features computed per vertex of the mesh, corresponding definitions, and dimensionality (i.e., number of channels $c$).}
    \begin{tabular}{l|l|r}
        \hline
        \textbf{Vertex feature} & \textbf{Definition} & \textit{c} \\
        \hline
        $\textbf{n}_i$ & Outward surface normal & 3 \\
        $g_i$ & Shortest geodesic to inlet & 1 \\
        $\gamma_i$ & Curvilinear abscissa & 1 \\
        $R_i$ & Radius of largest inscribed sphere & 1 \\
        $\textbf{T}_i$ & Tangent to centerline & 3 \\
        $\Phi_i$ & Laplacian eigenvectors & 6 \\
        $HKS_i$ & Heat kernel signature & 5 \\
        $WKS_i$ & Wave kernel signature & 5 \\
        $q_i$ & Blood inflow-rate & 1 \\
        $p_{in,i}$ & Pressure at coronary ostium & 1 \\
        $BC_{ID,i}$ & Node type in CFD analysis & 1 \\
        \hline
        & \textbf{Total input features} & \textbf{28} \\
        \hline
    \end{tabular}
    \label{tab:features}
\end{table}

\subsubsection{Network output}\label{sec_output}

We trained our models to predict scalar hemodynamic quantities on the domain $\mathcal{V}$, derived from steady blood flow through the coronary arteries. In experiments with the synthetic dataset, we conducted three groups of training, each targeting a different output variable: (\textit{i}) pressure $p_i$, (\textit{ii}) pressure drop - $\Delta p_i=p_{in}-p_i$ -, and (\textit{iii}) the vFFR field - $\mathrm{FFR}_i=p_i/p_{in}$. 
In experiments using patient-specific anatomies, we selected the pressure drop field as the output variable, as it achieved the best evaluation metrics in the synthetic dataset experiments.

\subsection{Evaluation metrics}

For each model, we reconstructed the corresponding vFFR field from the network output and evaluated its performance by comparing the predicted vFFR field $\hat{y}_i$ to the ground truth $y_i$ at each point. To quantify the similarity between the two fields, we used the per-point difference, defined as $\hat{y}_i - y_i$, and the approximation disparity \cite{suk2024deep}, defined as follows:

$$
\mathrm{Approx. \ Disp.} := \sum_{i=1}^N \left( \hat{y}_i - y_i \right) ^2 / \sum_{i=1}^N \hat{y}_i^2
$$

We also conducted a per-lesion analysis, comparing the predicted vFFR values in stenotic regions (specifically, the minimum vFFR value within each stenosis) against the ground truth for each model. This evaluation included correlation analysis, confusion matrix analysis, and Bland-Altman analysis. All statistical tests were conducted in Python using the SciPy library \cite{virtanen2020scipy}.

\subsection{Ablation studies}

We conducted ablation studies to assess the impact of input features on prediction accuracy. Specifically, we trained a single model using seven different combinations of input features and compared the resulting performance metrics. These experiments were conducted using the PointNet++ model, which offers a good trade-off between training time and performance and is a commonly used baseline in GDL tasks.

\section{Results} \label{results}

In this section, we present the results of our benchmark. First, we show the performance of each model on the synthetic dataset. Next, we present the results obtained on the patient-specific data. Finally, we provide insights from the ablation study, focusing on the influence of input features on training performance.

\subsection{Synthetic dataset}

\begin{figure}[h!]
\includegraphics[width=0.9\textwidth]{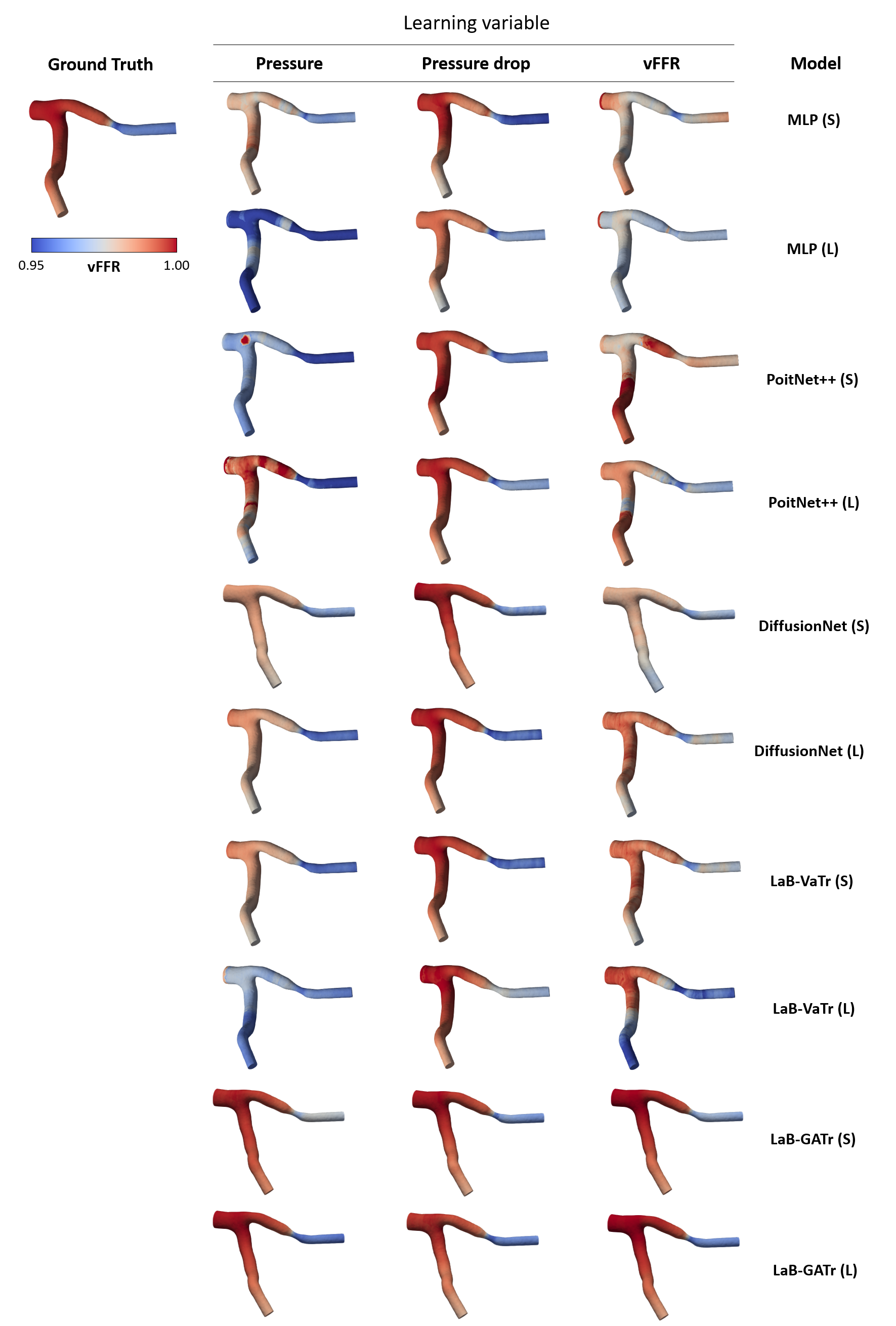}
\centering
\caption{Maps of vFFR computed from CFD (ground truth) on the synthetic bifurcation dataset, and inferred by each model, using different learning variables for one case of the test set.}
\label{fig:ffr_maps_syn}
\end{figure}

\begin{figure}[h]
\includegraphics[width=0.5\textwidth]{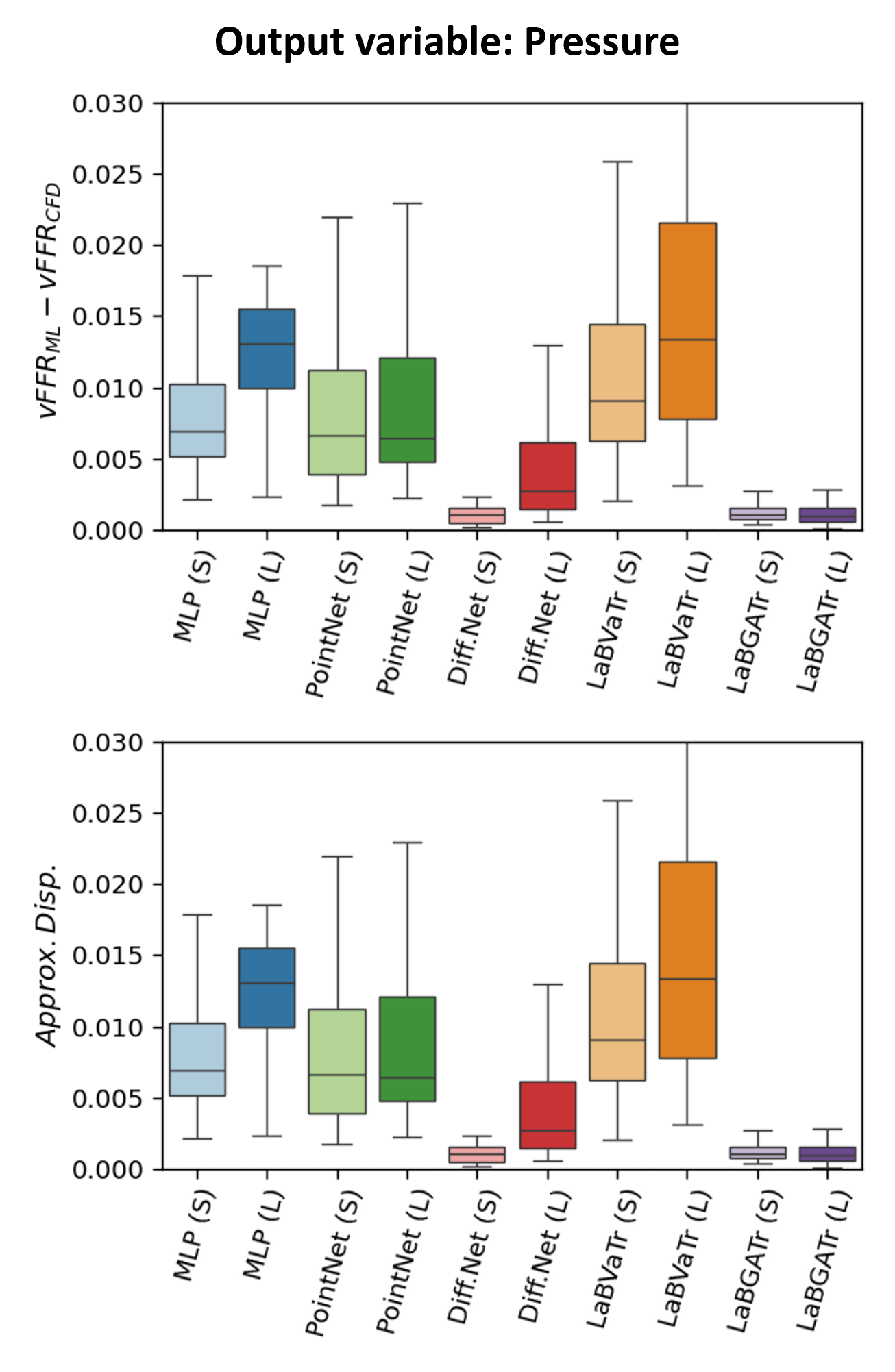}
\centering
\caption{Boxplots showing the performance of each model on the test set, learning pressure field, in terms of point-wise differences (top row) and approximation disparities (bottom row) between vFFR predictions and CFD-derived values.}
\label{fig:test_metrics_syn_a}
\end{figure}

Table \ref{tab:pressure_metrics} lists the performance of each model variant trained to predict the pressure field. During inference on the test set, all models tended to underestimate the per-point values. DiffusionNet and LaB-GATr were the only models to yield a per-point difference between the predicted and CFD-computed vFFR below $10^{-3}$. DiffusionNet and LaB-GATr also stood out in terms of approximation disparity. The boxplots for each metric are shown in Figure \ref{fig:test_metrics_syn_a}, while the inferred vFFR fields are illustrated in Figure \ref{fig:ffr_maps_syn}. Spearman's correlation and Bland-Altman analysis, conducted considering vFFR in stenotic regions, confirmed the superior performance of DiffusionNet and LaB-GATr.

\begin{table}[ht]
    \centering
    \caption{Synthetic dataset - Performance metrics for different models predicting pressure. Test loss is computed on the predicted pressure. Difference and approximation disparity are computed per point on the whole reconstructed vFFR field. Correlation coefficient and bias are computed in stenotic points. Training cost is expressed as seconds per epoch. For readability, standard deviations and limits of agreement are provided as supplementary material.}
    \resizebox{\textwidth}{!}{
    \begin{tabular}{lcccccc}
        \toprule
        \textbf{Network output} & \textit{Pressure} \\
        \midrule
        \textbf{Model} & \makecell{\textbf{Test Loss} \\ \textbf{(Pa $\times 10^{-2}$)} $\downarrow$} & 
        \makecell{\textbf{Per-point} \\ \textbf{Diff.} $\downarrow$} 
        & \makecell{\textbf{Approx.} \\ \textbf{Disp.} $\downarrow$} & $\boldsymbol{\rho}$ $\uparrow$ & \textbf{Bias} $\downarrow$ & \makecell{\textbf{Train. Cost} \\ \textbf{(s/epoch)} $\downarrow$} \\
        \midrule
        MLP (S)         & 9.90  & -8.10E-03 & 8.20E-03  & 0.658 & -0.0094 & \textbf{12}  \\
        MLP (L)         & 9.83  & -1.24E-02 & 1.25E-02  & 0.637 & -0.0171 & 17  \\
        PointNet++ (S)  & 9.25  & -1.02E-02 & 1.03E-02  & 0.519 & -0.0112 & 16  \\
        PointNet++ (L)  & 8.16  & -9.05E-03 & 9.22E-03  & 0.471 & -0.0093 & 18  \\
        DiffusionNet (S) & 5.46  & -7.10E-04 & \textbf{1.08E-03}  & 0.973 & -0.0008 & 28  \\
        DiffusionNet (L) & 5.29  & -4.60E-03 & 4.73E-03  & 0.698 & -0.0049 & 48  \\
        LaB-VaTr (S)    & 12.98 & -1.02E-02 & 1.06E-02  & 0.378 & -0.0105 & 35  \\
        LaB-VaTr (L)    & 11.89 & -1.44E-02 & 1.48E-02  & 0.253 & -0.0153 & 55  \\
        LaB-GATr (S)    & 4.54  & \textbf{-8.47E-04} & 1.23E-03  & 0.973 & \textbf{-0.0010} & 244 \\
        LaB-GATr (L)    & \textbf{3.37}  & -8.83E-04 & 1.12E-03  & \textbf{0.976} & -0.0010 & 419 \\
        \bottomrule
    \end{tabular}}
    \label{tab:pressure_metrics}
\end{table}

\begin{figure}[h]
\includegraphics[width=0.5\textwidth]{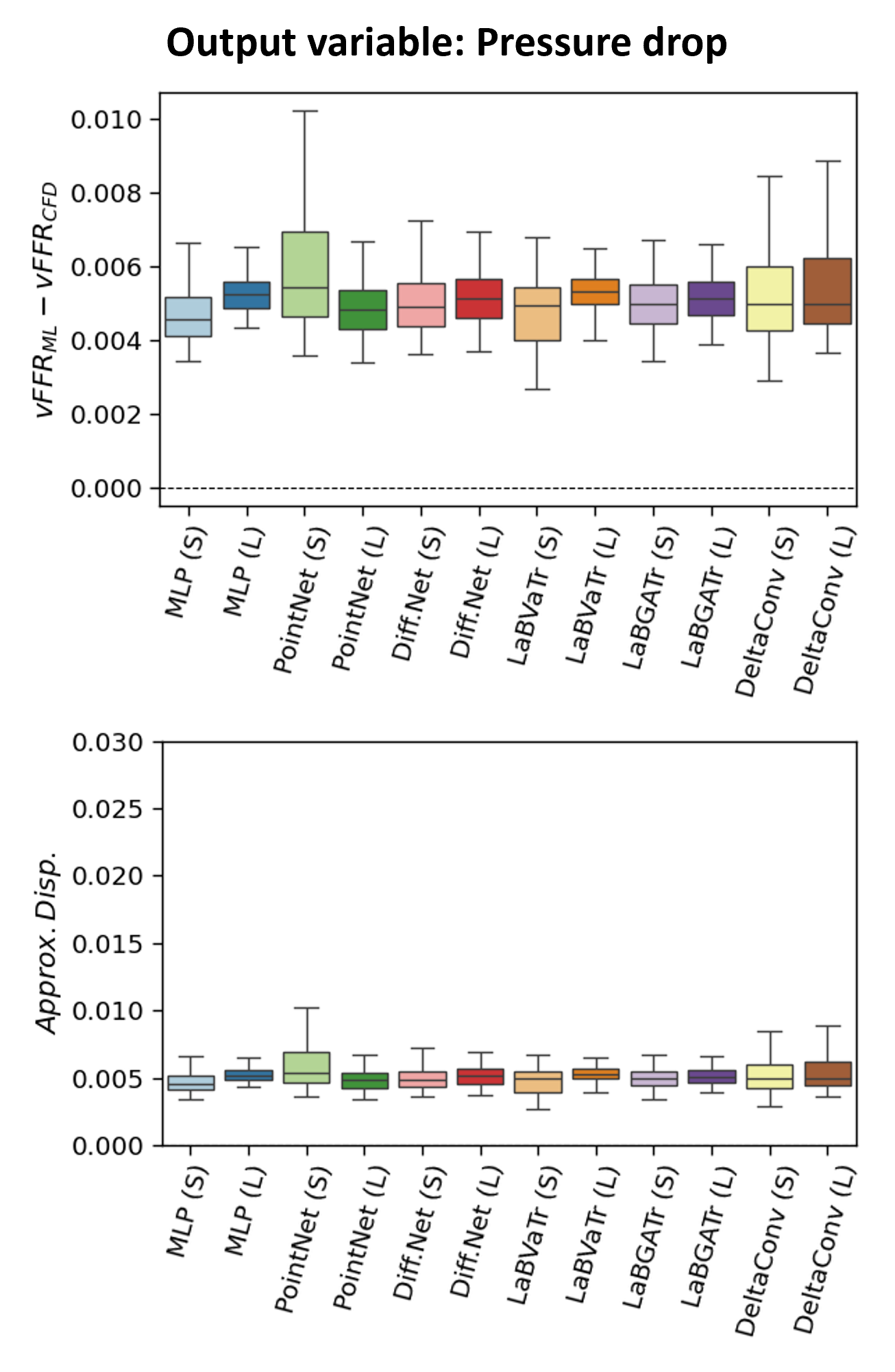}
\centering
\caption{Boxplots showing the performance of each model on the test set, learning pressure drop field, in terms of point-wise differences (top row) and approximation disparities (bottom row) between vFFR predictions and CFD-derived values.}
\label{fig:test_metrics_syn_b}
\end{figure}

Similarly, Table \ref{tab:pressure_drop_metrics} summarizes the results of training with the standardized pressure drop field as the network output. In inference, the per-point difference and approximation disparity were comparable across all models, averaging approximately $\sim$4.11E-3 and $\sim$5.1E-3, respectively (Figure \ref{fig:test_metrics_syn_b}). The reconstructed vFFR map derived from the inferred pressure drop field was qualitatively the most accurate (Figure \ref{fig:ffr_maps_syn}). Notably, the pressure drop was also the only output variable for which we were able to stably train the DeltaConv with good performance. All models demonstrated strong ($>0.4$) to very strong ($>0.7$) correlation between the inferred and CFD-based vFFR. The bias was consistent across models, with LaB-GATr exhibiting the narrowest 95\% limit of agreement band (see Supplementary).

\begin{table}[htbp]
    \centering
    \caption{Synthetic dataset - Performance metrics for different models predicting pressure drop. Test loss is computed on the predicted pressure drop. Difference and approximation disparity are computed per point on the whole reconstructed vFFR field. Correlation coefficient and bias are computed in stenotic points. Training cost is expressed as seconds per epoch. For readability, standard deviations and limits of agreement are provided as supplementary material.}
    \resizebox{\textwidth}{!}{
    \begin{tabular}{lcccccc}
        \toprule
        \textbf{Network output} & \textit{Pressure drop} \\
        \midrule
        \textbf{Model} & \makecell{\textbf{Test Loss} \\ \textbf{(Pa $\times 10^{-2}$)} $\downarrow$} & 
        \makecell{\textbf{Per-point} \\ \textbf{Diff.} $\downarrow$} 
        & \makecell{\textbf{Approx.} \\ \textbf{Disp.} $\downarrow$} & $\boldsymbol{\rho}$ $\uparrow$ & \textbf{Bias} $\downarrow$ & \makecell{\textbf{Train. Cost} \\ \textbf{(s/epoch)} $\downarrow$} \\
        \midrule
        MLP (S)         & 7.50 & \textbf{3.23E-03} & \textbf{4.69E-03} & 0.940 & 0.0021 & \textbf{12}  \\
        MLP (L)         & 9.60 & 4.15E-03 & 5.25E-03 & 0.967 & 0.0021 & 17  \\
        PointNet++ (S)  & 8.10 & 3.93E-03 & 5.81E-03 & 0.976 & 0.0022 & 16  \\
        PointNet++ (L)  & 7.08 & 3.98E-03 & 4.84E-03 & 0.977 & \textbf{0.0021} & 18  \\
        DiffusionNet (S) & 6.84 & 3.54E-03 & 5.06E-03 & 0.971 & 0.0022 & 28  \\
        DiffusionNet (L) & 4.69 & 3.91E-03 & 5.14E-03 & \textbf{0.986} & 0.0024 & 48  \\
        LaB-VaTr (S)    & 8.58 & 4.13E-03 & 4.72E-03 & 0.935 & 0.0026 & 35  \\
        LaB-VaTr (L)    & 11.58 & 4.59E-03 & 5.32E-03 & 0.791 & 0.0033 & 55  \\
        LaB-GATr (S)    & 6.34 & 4.39E-03 & 4.97E-03 & 0.979 & 0.0033 & 244 \\
        LaB-GATr (L)    & \textbf{3.78} & 4.19E-03 & 5.12E-03 & 0.967 & 0.0030 & 419 \\
        DeltaConv (S)    & 11.23 & 4.47E-03 & 5.28E-03 & 0.562 & 0.0021 & 17  \\
        DeltaConv (L)    & 11.56 & 4.84E-03 & 5.42E-03 & 0.651 & 0.0026 & 30  \\
        \bottomrule
    \end{tabular}}    \label{tab:pressure_drop_metrics}
\end{table}

\begin{figure}[h]
\includegraphics[width=0.5\textwidth]{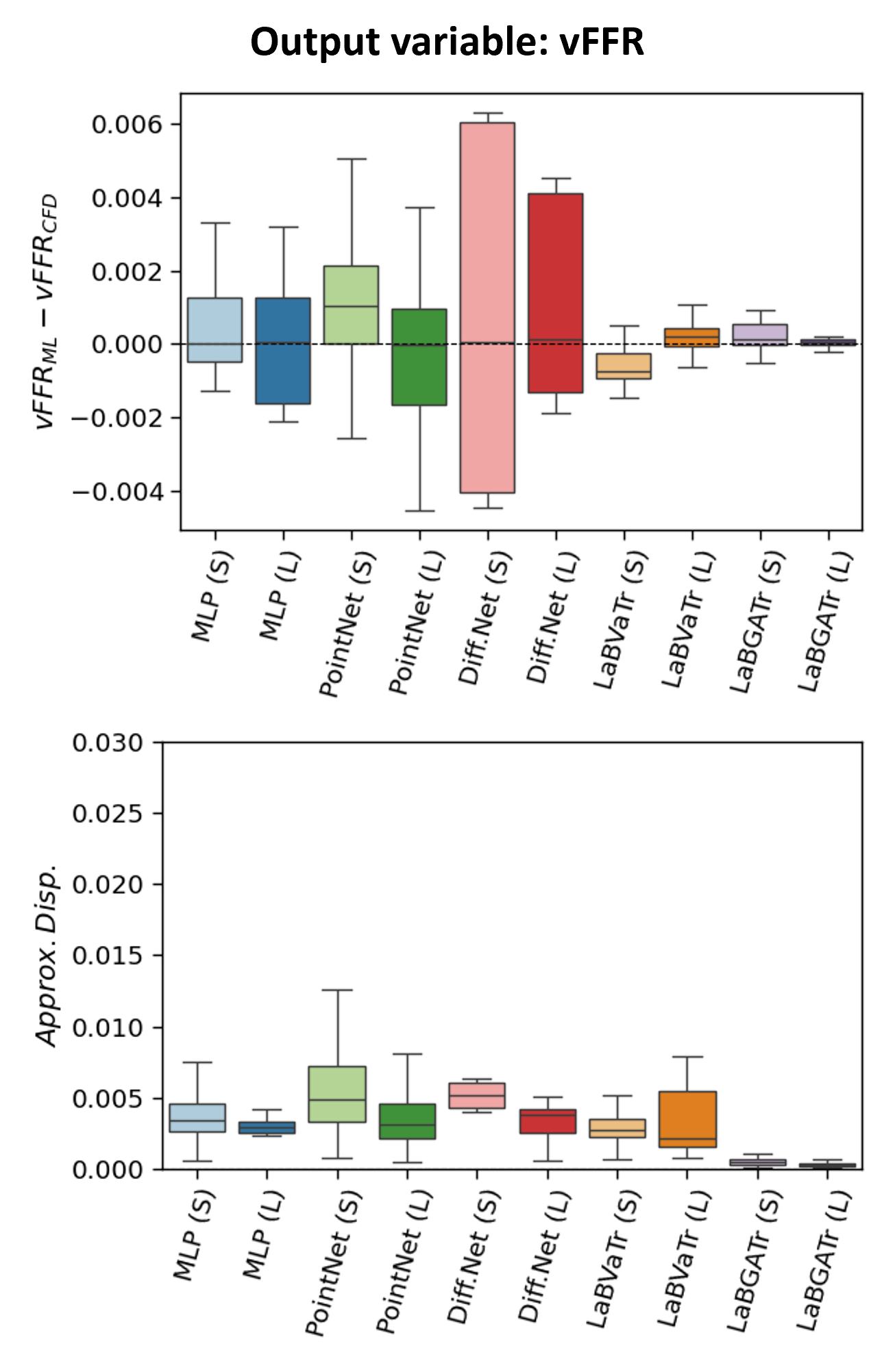}
\centering
\caption{Boxplots showing the performance of each model on the test set, learning vFFR field, in terms of point-wise differences (top row) and approximation disparities (bottom row) between vFFR predictions and CFD-derived values.}
\label{fig:test_metrics_syn_c}
\end{figure}

Finally, Table \ref{tab:vffr_metrics} illustrates the results achieved in the experiments using the vFFR field directly as the output variable. In inference, LaB-GATr outperformed all the tested model, yielding a per-point difference of 5.35E-05 and an approximation disparity of 5.40E-04 (Figure \ref{fig:test_metrics_syn_c}). In particular, LaB-GATr was the only model to predict a vFFR field that qualitatively resembles the ground truth (Figure \ref{fig:ffr_maps_syn}). The MLP demonstrated comparable test metrics to the LaB-GATr. However, in the statistical analysis of stenotic regions, the MLP showed a lower correlation coefficient and a bias that was an order of magnitude larger.

\begin{table}[htbp]
    \centering
    \caption{Synthetic dataset - Performance metrics for different models predicting vFFR. Test loss, difference and approximation disparity are computed per point on the whole reconstructed vFFR field. Correlation coefficient and bias are computed in stenotic points. Training cost is expressed as seconds per epoch. For readability, standard deviations and limits of agreement are provided as supplementary material.}
    \resizebox{\textwidth}{!}{
    \begin{tabular}{lcccccc}
        \toprule
        \textbf{Network output} & \textit{vFFR} \\
        \midrule
        \textbf{Model} & \makecell{\textbf{Test Loss} \\ \textbf{($\times 10^{-3}$)} $\downarrow$} & 
        \makecell{\textbf{Per-point} \\ \textbf{Diff.} $\downarrow$} 
        & \makecell{\textbf{Approx.} \\ \textbf{Disp.} $\downarrow$} & $\boldsymbol{\rho}$ $\uparrow$ & \textbf{Bias} $\downarrow$ & \makecell{\textbf{Train. Cost} \\ \textbf{(s/epoch)} $\downarrow$} \\
        \midrule
        MLP (S)         & 3.63 & 3.89E-04  & 3.66E-03 & 0.287 & 0.0016  & \textbf{12}  \\
        MLP (L)         & 2.66 & -5.76E-05 & 2.67E-03 & 0.650 & 0.0012  & 17  \\
        PointNet++ (S)  & 4.69 & 1.08E-03  & 5.59E-03 & 0.472 & 0.0015  & 16  \\
        PointNet++ (L)  & 2.94 & -3.21E-04 & 3.62E-03 & 0.743 & -0.0004 & 18  \\
        DiffusionNet (S) & 4.65 & 8.21E-04  & 4.67E-03 & 0.630 & 0.0009  & 28  \\
        DiffusionNet (L) & 3.23 & 1.19E-03  & 3.25E-03 & 0.726 & 0.0011  & 48  \\
        LaB-VaTr (S)    & 2.76 & -5.78E-04 & 2.79E-03 & 0.616 & -0.0016 & 35  \\
        LaB-VaTr (L)    & 3.28 & 2.02E-04  & 3.32E-03 & 0.643 & -0.0017 & 55  \\
        LaB-GATr (S)    & 0.53 & 2.28E-04  & 5.40E-04 & 0.970 & 0.0004  & 244 \\
        LaB-GATr (L)    & \textbf{0.31} & \textbf{5.35E-05}  & \textbf{3.13E-04} & \textbf{0.988} & \textbf{-0.0002} & 419 \\
        \bottomrule
    \end{tabular}}
    \label{tab:vffr_metrics}
\end{table}

\subsection{Patient-specific dataset}
\begin{figure}[h!]
\includegraphics[width=\textwidth, angle=0]{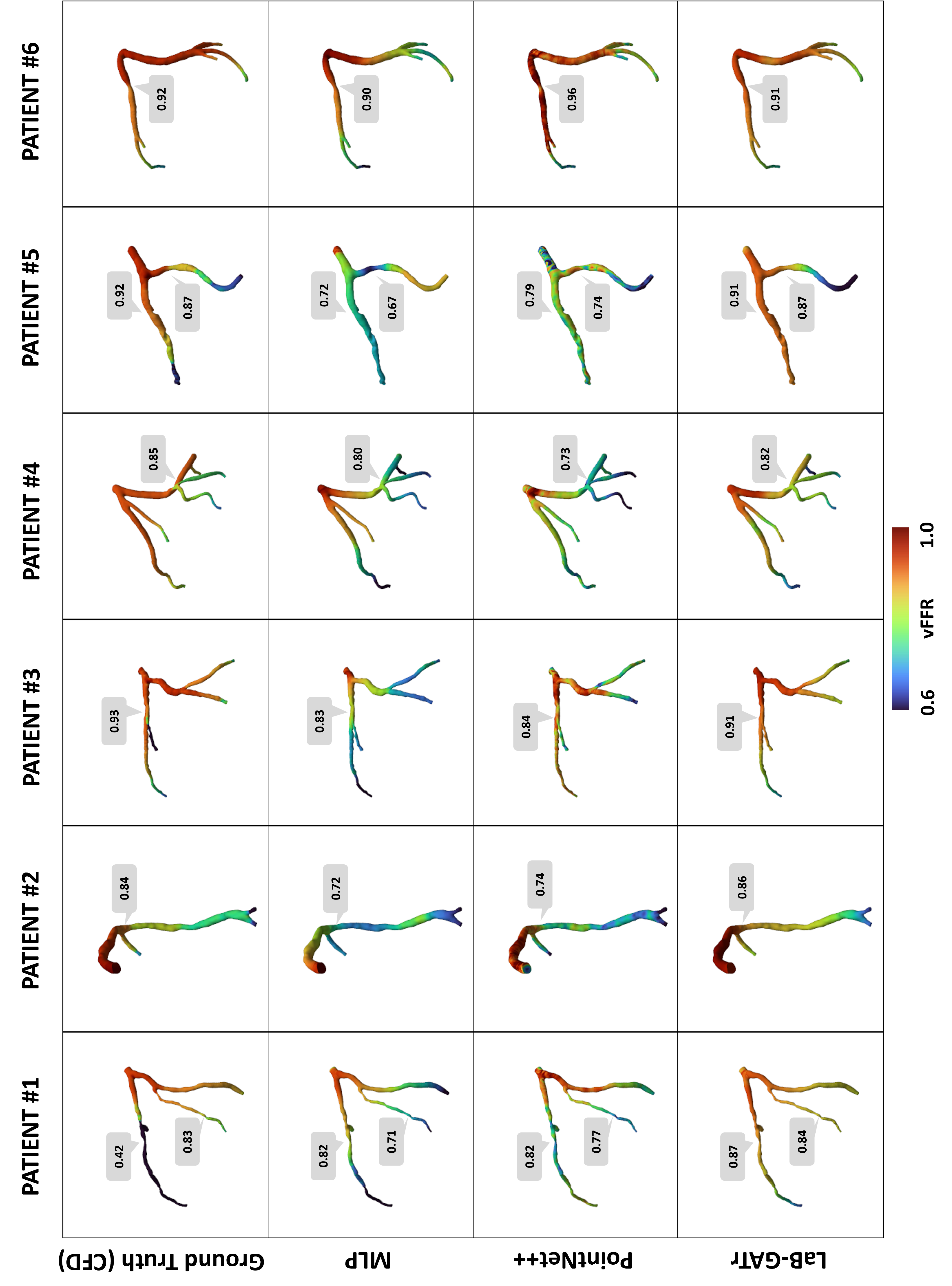}
\centering
\caption{Maps of vFFR computed from CFD (ground truth) on the patient' specific dataset, and inferred by three representative models (MLP (S), PointNet++ (S) and LaB-GATr (S)) on 6 test set cases, using pressure drop as learning variables. The vFFR value is indicated in correspondence with lesion sites.}
\label{fig:ffr_maps_ps}
\end{figure}

In this section, we present the results of training our models on a dataset of patient-specific CFD analyses. Overall, all models demonstrated reduced performance during inference compared to the results on synthetic anatomies. Among the models, LaB-GATr consistently showed the best generalization capability, particularly in its 'small' variant. Most models correctly predicted the pressure gradient direction from the inlet to the outlet; however, some failed to capture sharp pressure drops in stenotic regions. 
Figure \ref{fig:ffr_maps_ps} shows for each model (rows) the reconstructed vFFR field for 6 cases (columns) of the test set. We observed moderate agreement between the inferred vFFR and the ground truth for \textit{Patients 2}, \textit{3}, \textit{4}, and \textit{6}. For \textit{Patient 5}, only the S-variant of LaB-GATr accurately predicted the vFFR gradient in the circumflex artery, while for \textit{Patient 1}, none of the models successfully captured the vFFR drop in the stenosis of the left anterior descending artery. 
In general, all models showed greater discrepancies for meshes with numerous bifurcations in sub-branches. Conversely, for cases involving a single main vessel (e.g., right coronary arteries), the models demonstrated better agreement with the ground truth.
Table \ref{tab:ps_metrics_part1} and \ref{tab:ps_metrics_part2} report the performance metrics for each model. All models achieved a validation loss of $\sim10$ to 15 Pa. In front of a significantly higher computational cost, LaB-GATr was the model that again yielded best results. DiffusionNet (S-variant) yielded the highest correlation coefficient (0.72), and together with LaB-GATr (S-variant) was the only model to achieve an accuracy above 0.7; however it poorly performed in terms of precision and exhibited a larger bias compared to LaB-GATr.

\begin{table}[htbp]
    \centering
\caption{Patient-specific dataset - Performance metrics for various models. Test loss, training cost, per-point difference, and approximation disparity are computed pointwisely on the whole mesh. The test loss is computed on the predicted pressure drop and expressed in Pascal. \textbf{Bold} font indicates the best result for each metric.}
\begin{tabular}{lcccc}
        \toprule
        \textbf{Model} & \makecell{\textbf{Test Loss} \\ \textbf{(Pa)} $\downarrow$} & \makecell{\textbf{Per-point} \\ \textbf{Diff.} $\downarrow$} & \makecell{\textbf{Approx.} \\ \textbf{Disp.} $\downarrow$} & \makecell{\textbf{Train. Cost} \\ \textbf{(s/epoch)} $\downarrow$} \\
        \midrule
        MLP (S)         & 13.6±7.6 & -0.11±0.12 & 0.172±0.15 & 23 \\
        MLP (L)         & 13.0±7.0 & -0.10±0.11 & 0.169±0.15 & 24 \\
        DiffusionNet (S) & 15.3±8.3 & -0.06±0.13 & 0.149±0.20 & 44 \\
        DiffusionNet (L) & 14.7±8.9 & -0.09±0.14 & 0.181±0.19 & 83 \\
        PointNet++ (S)   & 15.1±8.9 & -0.09±0.14 & 0.181±0.20 & \textbf{17} \\
        PointNet++ (L)   & 14.0±8.4 & -0.09±0.13 & 0.170±0.19 & 18 \\
        LaB-VaTr (S)     & 14.5±7.5 & -0.10±0.12 & 0.180±0.19 & 42 \\
        LaB-VaTr (L)     & 13.5±6.9 & -0.11±0.11 & 0.165±0.14 & 54 \\
        LaB-GATr (S)     & \textbf{11.4±6.5} & \textbf{-0.03±0.10} & 0.123±0.12 & 104 \\
        LaB-GATr (L)     & 12.0±8.2 & -0.04±0.13 & \textbf{0.116±0.14} & 154 \\
        \bottomrule
    \end{tabular}
    \label{tab:ps_metrics_part1}
\end{table}

\begin{table}[htbp]
    \centering
    \caption{Patient-specific dataset - Performance metrics for various models. Bias [LoA=Limits of Agreement], correlation coefficient ($\rho$), precision (Prec.), recall (Rec.), accuracy (Acc.), and limits of agreement (LoA) are calculated on vFFR predictions in stenotic regions. \textbf{Bold} font indicates the best result for each metric.}
    \begin{tabular}{lcccccc}
        \toprule
        \textbf{Model} & $\boldsymbol{\rho}$ $\uparrow$ & \textbf{Bias} $\downarrow$ & \textbf{LoA} & \textbf{Prec.} $\uparrow$ & \textbf{Rec.} $\uparrow$ & \textbf{Acc.} $\uparrow$ \\
        \midrule
        MLP (S)         & 0.58 & -0.077 & [-0.30; 0.15] & 0.40 & 0.91 & 0.58 \\
        MLP (L)         & 0.66 & -0.082 & [-0.29; 0.13] & 0.41 & \textbf{1.00} & 0.58 \\
        DiffusionNet (S) & \textbf{0.72} & -0.054 & [-0.27; 0.17] & 0.50 & 0.91 & 0.71 \\
        DiffusionNet (L) & 0.68 & -0.060 & [-0.29; 0.18] & 0.46 & 0.91 & 0.66 \\
        PointNet++ (S)   & 0.53 & -0.051 & [-0.31; 0.20] & 0.46 & \textbf{1.00} & 0.66 \\
        PointNet++ (L)   & 0.67 & -0.059 & [-0.29; 0.17] & 0.46 & \textbf{1.00} & 0.66 \\
        LaB-VaTr (S)     & 0.68 & -0.080 & [-0.31; 0.14] & 0.42 & \textbf{1.00} & 0.61 \\
        LaB-VaTr (L)     & 0.66 & -0.087 & [-0.32; 0.14] & 0.41 & \textbf{1.00} & 0.58 \\
        LaB-GATr (S)     & 0.69 & \textbf{0.004} & [-0.20; 0.14] & \textbf{0.67} & 0.55 & \textbf{0.79} \\
        LaB-GATr (L)     & 0.69 & -0.033 & [-0.21; 0.22] & 0.53 & 0.73 & 0.68 \\
        \bottomrule
    \end{tabular}
    \label{tab:ps_metrics_part2}
\end{table}

\subsection{Impact of different input features}
Table \ref{tab:ablation} summarizes the results of the ablation study conducted on input features, using PointNet++ as the backend and trained on the synthetic dataset of bifurcations. Our findings highlight that specific features, such as surface normals ($\textbf{n}_i$), geodesics ($g_i$), and flow rate ($q_i$), are critical for achieving strong model performance, while features derived from Laplacian eigenvectors provide only modest improvements. Including the inlet pressure ($p_{in,i}$) as an input feature significantly enhances the predictive capability of the model and correctly sets the inferred pressure range. Local radius ($R_i$) benefits the generalization capability of the model as well. Interestingly, the model yielded the best results when using either all the features from this study or a limited subset of those features (\textit{Ablation 3}), which included few local geometric and physical descriptors.

\begin{table}[h!]
\caption{Comparison of different feature combinations (indicated as \textit{Ablation n°}) for predicting the pressure drop field. Features are indicated with symbols introduced in Table \ref{tab:features}. Pointwise difference is reported as evaluation metric. \textbf{Bold} values indicate the two best results achieved.}
\centering
\resizebox{\textwidth}{!}{
\begin{tabular}{|c|ccccccccccc|c|}

\hline
 & \textbf{n}\_i & $g$\_i & $\gamma$\_i & $R$\_i & $\Phi$\_i & HKS\_i & WKS\_i & \textbf{T}\_i & BC\_{ID,i} & $q$\_i & $p\_{in, i}$ & \makecell{\textbf{Pointwise} \\ \textbf{Diff. (Pa $\times 10^{-2}$)} $\downarrow$ } \\ \hline
\textit{Ablation 1} & \tick & \tick & \cross & \cross & \cross & \cross & \cross & \cross & \cross & \tick & \cross & 10.56 \\
\textit{Ablation 2} & \tick & \tick & \cross & \tick & \cross & \cross & \cross & \cross & \cross & \tick & \cross & 7.92 \\
\textit{Ablation 3} & \tick & \tick & \cross & \tick & \cross & \cross & \cross & \cross & \cross & \tick & \tick & \textbf{6.96} \\
\textit{Ablation 4} & \tick & \tick & \cross & \cross & \cross & \cross & \cross & \cross & \cross & \tick & \tick & 8.10 \\
\textit{Ablation 5} & \tick & \tick & \tick & \tick & \tick & \tick & \tick & \cross & \cross & \tick & \cross & 7.50 \\
\textit{Ablation 6} & \tick & \tick & \tick & \tick & \cross & \cross & \cross & \tick & \cross & \tick & \tick & 8.34 \\
\textit{Ablation 7} & \cross & \cross & \cross & \cross & \tick & \tick & \tick & \tick & \cross & \tick & \tick & 10.68 \\
\textit{All features}   & \tick & \tick & \tick & \tick & \tick & \tick & \tick & \tick & \tick & \tick & \tick & \textbf{7.08} \\ \hline

\end{tabular}}
\label{tab:ablation}
\end{table}

\section{Discussion} \label{discuss}
In this work, we have presented a systematic benchmark between different GDL models to predict scalar fields derived from the solution of steady CFD analysis on both synthetic and real-life patient anatomies of coronary vessels. While the 3D mesh of the artery can be easily reconstructed from CT, the measurement of functional indices used for the assessment of stenosis severity, such as FFR, requires invasive examination or time-consuming CFD simulations, which are both economically expensive solutions. We conditioned the tested models on geometric features that can be computed directly from the mesh and physical quantities (e.g., flow rate, pressure) that can be directly or indirectly derived from clinical measurements, usually taken during patient screening. We first evaluated each model on a dataset of CFD simulations run on 1500 synthetic anatomies, inferring different outputs from which the vFFR field was then reconstructed. Then, we tested each model on a dataset of 427 CFD simulations run on patient-specific anatomies, inferring the pressure drop, that is, the output variable that yielded the best performances on the synthetic dataset. 
Pre-computing the input features to inform each model required approximately 1 minute on average, while during inference, all models took less than 30 seconds to produce the output field at hand, limiting the overall cost of the ML-based estimation of vFFR to under 2 minutes.

\subsection{Impact of network output}
In our experiments on the synthetic dataset, we observed that all models generally had the best performance, both qualitative (Figure \ref{fig:ffr_maps_syn}) and quantitative (Table \ref{tab:pressure_drop_metrics}) when inferring the pressure drop field. We suggest that this is due to the fact that, while the pressure field can range in very heterogeneous intervals across arteries depending both on the pressure drop along the vessel and the pressure at the inlet (i.e., patient's systemic pressure), the pressure drop field is generally more homogeneous across patients, as it always starts from zero at the ostium and is more correlated with geometric features, such as vessel radius and length, especially under Poiseuille's flow hypothesis. 
When training the models on the pressure drop, we found a low common bias of $\sim 4$E-03 on the derived vFFR field, while training on the pressure produced for all the models a negative bias (as visible in Figure \ref{fig:test_metrics_syn_b}). This could be due to a common underestimation of the output variable for both cases, which in the case of inference on the pressure field results in a lower distal pressure ($P_d$), and thus a lower vFFR\textsubscript{ML} - vFFR\textsubscript{CFD}, as it is defined as $(P_{d,ML}-P_{d,CFD}) / P_{in}$. On the other hand, in the case of the pressure drop, underestimating the output leads to systematically overestimating the vFFR\textsubscript{ML}, defined as $(P_{in}-\Delta P_{ML})/P_{in}$. Inferring vFFR directly, produced no systematic bias, but yielded a significantly lower performance, as shown in Figure \ref{fig:test_metrics_syn_c} by the wider percentile ranges.

Training on the pressure drop resulted, in general, in more stable training runs; in particular, this is the only network output for which DeltaConv training was stable, achieving good performance metrics. We trained DeltaConv on the other network variable as well, however the performance achieved were not as good as training on the pressure drop, using the architecture implementation as presented in Ref. \cite{wiersma2022deltaconv}. Given that the network was successfully trained for one output variable, we expect that with proper adjustment to the backend implementation it is feasible to train it on the other learning variable as well, however this goes beyond the purpose of this work, still hinting a possible development.

\subsection{Prediction of vFFR in real-life patients}
When inferring the pressure drop onto patient specific data, we tested both the capability to correctly infer pressure on the whole mesh and the diagnostic accuracy in predicting the vFFR in stenotic regions. In general, all models predicted a smooth vFFR decrease going from the inlet to outlets, but LaB-GATr was the only one that accurately permitted to reconstruct the vFFR field, capturing sharper decreases due to stenosis. On average, LaB-GATr yielded an error in the order of 1 to 10 Pa on the predicted pressure drop field (and thus on pressure). Previous works have proved the capability of analogous model to learn fluid dynamic fields in vessel-like shaped domains. Rygiel \textit{et al.} \cite{rygiel2023centerlinepointnet++} obtained excellent agreement between CFD-based vFFR and machine learning prediction. Similarly, Suk \textit{et al.} \cite{suk2024mesh, suk2023se} proved the capability of the model to learn wall shear stress and velocity field. Nevertheless, in all these studies, experiments were conducted on synthetically generated anatomies, which simplify the task compared to real-life geometries that exhibit significant heterogeneity and complexity. 

Two recent studies achieved patient-specific hemodynamic predictions using deep learning. Yang \textit{et al.} \cite{yang2024attention} proposed a novel attention-based \cite{vaswani2017attention} variant of the popular U-Net architecture \cite{ronneberger2015u} to compute vFFR in patient-specific coronary artery anatomies. They trained the model on synthetically generated vessel segments and inferred the vFFR on stenotic segments of patient coronary arteries, thus overcoming the limitation of poor availability of patient-specific data. Pegolotti \textit{et al.} \cite{pegolotti2024learning} achieved good performance in predicting flow rate and pressure in patient-specific anatomies by training their network on 1D flow solutions. While both studies successfully predicted hemodynamics in patient-specific anatomies, they introduced significant simplifications, such as reducing the volume of interest to small vessel segments and relying on reduced-order 1D solutions of CFD analyses. 

Our benchmark, along with results reported by other authors in the literature, supports the notion that inferring scalar fields on simple topologies (e.g., 2D manifolds representing small portions of anatomy, synthetic data, or 1D manifolds) is a task manageable by many deep learning architectures. However, when the topology becomes more complex and exhibits high variability within the dataset, as in the case of patient-specific data, transformer-based architectures emerge as the most suitable option for addressing the task effectively, both due to the efficient global context aggregation and the straight-forward processing of shapes. 

\subsection{Impact of different input features}
Our ablation study, conducted to identify the crucial parameters for informing the model and achieving optimal performance, highlighted the importance of certain inputs in granting the model generalization capability. Specifically, geometric parameters such as surface normals and geodesic distance to artery landmarks (intrinsic orientation) proved essential for accurate predictions. When these parameters (\textit{Ablation 7}) were removed from the input features, the evaluation metric (i.e., the vFFR pointwise difference) worsened by 2\% to 59\% compared to any other tested combination in which they were included. Similarly, the local vessel radius (\textit{Ablation 2} vs. \textit{Ablation 1}) and the pressure at the inlet (\textit{Ablation 3} vs. \textit{Ablation 2}) consistently improved model performance.
Providing inlet pressure can be interpreted as setting a Dirichlet condition at the inlet, effectively constraining a subset of the mesh vertices to assume specific values. As a result, the model's predictions were generally more accurate near the inlet nodes and deteriorated as the distance from the ostium increased (as visible in Figure \ref{fig:ffr_maps_syn} for the MLP (L) model using vFFR as the learning variable). Local descriptors derived from the spectral decomposition of the Laplacian matrix enhanced the model's performance but, when used as the sole descriptors of the anatomy, produced the poorest results (\textit{Ablation 7}). Overall, we found that a small subset of input features (\textit{Ablation 3}) is sufficient to learn the target scalar field, achieving comparable results to those obtained with the full feature set considered in this study.

For the ablation study, we used PointNet++ as the backend, as it had the lowest training cost in the benchmark and is widely adopted for GDL tasks. Given its efficiency and versatility, PointNet++ could be employed as a preliminary step in future experiments to perform feature selection and identify optimal input combinations.

\subsection{Limitations}
The presented work is not exempt from limitations. Firstly, systematically comparing different backends is challenging due to their varying implementations. We selected the number of network parameters as the primary criterion and mitigated bias from specific model sizes by defining two versions for each model. However, this approach could not be applied to LaB-GATr, as its parameter space is inherently constrained compared to typical GDL models. Moreover, training LaB-GATr with a parameter count comparable to the models tested in this work ($\sim$1M to 4M parameters) is infeasible due to its high computational cost. Another limitation is the size of the dataset of patient-specific anatomies. While a dataset containing 427 samples might be sufficiently large for other tasks, the significant variability, both in the topology of the manifolds and in the scalar field we aim to infer, renders this dataset insufficient to represent all the cases required for robust model generalization. Finally, we cannot definitively conclude that the pressure drop, which yielded the best results as the network's output variable in experiments on the synthetic dataset, exhibits the same behavior when applied to real-life patient anatomies; even though evidence in the literature \cite{rygiel2023centerlinepointnet++, suk2024deep} suggests that pressure drop is indeed an optimal learning variable for this type of task.

\section{Conclusion} \label{conclusion}
We have presented a benchmark of different model backends, comparing their generalization capabilities in learning scalar fields derived from the solutions of CFD analyses on the surface meshes of coronary arteries. Our results indicate that inferring the pressure drop field on topologically simple anatomies (i.e., a bifurcation) is a task that can be successfully handled by various commonly used geometric deep learning models. Conversely, when predicting pressure drop on real-life patient geometries, only LaB-GATr demonstrated robust generalization capabilities, achieving the best evaluation metrics and diagnostic performance on stenotic segments. To achieve high performance, the model must be provided with essential local geometric and physical descriptors. Our findings indicate that surface normals, geodesics, vessel radius, inflow, and inlet pressure are crucial for this task.

In conclusion, various geometric deep learning models have demonstrated potential as CFD surrogates for FFR estimation when applied to problems involving relatively simple geometries (e.g., a straight vessel or a bifurcation). However, when considering patient-specific anatomies, we only find transformer-based backends (i.e., LaB-GATr) capable of learning the solutions of steady-state CFD analyses. This capability positions them as effective surrogates for accelerating the diagnostic process of CAD by directly predicting hemodynamic parameters.

\section*{Acknowledgments}

PNRR-POC-2022-12376500 Financed by the Ministry of Health funds PNRR M6C2I2.1 and by the European Union – NextGenerationEU 

This work was partly funded by the National Plan for NRRP Complementary Investments (PNC), established with the decree-law 6 May 2021, n. 59, converted by law n. 101 of 2021) in the call for the funding of research initiatives for technologies and innovative trajectories in the health and care sectors (Directorial Decree n. 931 of 06-06-2022) - project n. PNC0000003 - AdvaNced Technologies for Human-centrEd Medicine (project acronym: ANTHEM). This work reflects only the authors’ views and opinions, neither the Ministry for University and Research nor the European Commission can be considered responsible for them.

This project has received funding from the European Union’s Horizon Europe research and innovation programme under grant agreement No 101080947 (VASCUL-AID). 

The authors would like to acknowledge Ruben Wiersma, PhD for his implementation and for guidance on the use of the DeltaConv architecture, which has been adopted in this work (https://github.com/rubenwiersma/deltaconv).

\bibliographystyle{elsarticle-num} 
\bibliography{bibliography}

\end{document}